\documentclass[manuscript]{aastex}

\begin{document}
\pagenumbering{arabic}

\title{COMPARISON BETWEEN THE GLOBULAR CLUSTERS IN NGC 5128 AND THE GALAXY GLOBULAR}

\author{Sidney van den Bergh}
\affil{Dominion Astrophysical Observatory, Herzberg Institute of Astrophysics, National Research Council of Canada, 5071 West Saanich Road, Victoria, BC, V9E 2E7, Canada}
\email{sidney.vandenbergh@nrc.gc.ca}

\begin{abstract}

Some of the properties of the globular clusters in
NGC 5128 - the nearest giant elliptical galaxy - are compared to those of Galactic globular clusters. Assuming the color- metallicity relations that hold for Galactic globular clusters one finds that the metal-poor clusters in NGC 5128 that have [Fe/H] $<$ -1.80 are significantly fainter than are the more metal-rich globulars in that galaxy. No such metallicity dependent luminosity difference is observed among the globular clusters associated with the Milky Way System.

Furthermore the NGC 5128 cluster sample contains two objects that, on the basis of their observed colors appear to be super metal-poor. It is speculated that many of these apparently faint and metal-poor clusters in NGC 5128 are actually objects resembling intermediate-age Galactic open clusters.

It is also found that large clusters with FWHM $>$ 10 pc are
typically less luminous in NGC 5128 than are their more more compact counterparts. In this respect the NGC 5128 cluster system is similar to the Galactic globular cluster system, in which large clusters also tend to be less luminous than compact ones. Finally, the present data may hint at the possibility that the NGC 5128 cluster system differs from that surrounding the Milky Way, in the sense that the NGC 5128 objects do not seem to exhibit a clear cut gap between the regions of the FWHM vs $M_{v}$ plane that are occupied by globular clusters and by dwarf spheroidal galaxies.

\end{abstract}


\section{INTRODUCTION}

NGC 5128 (Centaurus A), which is located at a distance of only
$\sim$4 Mpc, is the nearest giant elliptical galaxy. Its proximity facilitates comparison of some of the properties of its globular cluster system with those of the globulars associated with the Milky Way System. Such a comparison is of particular interest because it might throw some light on possible differences in the early evolutionary histories of massive spiral and elliptical galaxies. An obvious caveat is that two galaxies constitute a small sample, and that they might turn out not be representative of the the class of massive ellipticals and of massive spiral galaxies. It would therefore be important to obtain observations that could allow one extend the present work to cluster systems associated with other massive elliptical and spiral galaxies.
The globular cluster system surrounding the nearby galaxy M31 has not been included in the present investigation because the quality of the color, diameter, reddening and metallicity data for many of the clusters in the Andromeda galaxy is still quite low. Inclusion of such data might therefore confuse, rather than clarify, some of the issues discussed in the present paper.

\section{GLOBULAR CLUSTER DATA}

 Data on the luminosities, diameters and metallicities of 115
putative globular clusters in NGC 5128 were taken from information published by Harris et al. (2006) and is listed in Table 1. The highly metallicity sensitive index $C - T_{1}$ on the Washington photometric system was used to assign [Fe/H] values to individual clusters. Cluster luminosities were derived by assuming that the turnover luminosity $T_{1}$ = 20.35 of the globulars in NGC 5128 corresponds to $M_{v} \simeq$ -7.4 (Harris 2001). Furthermore Geisler \& Forte
(1990) find E(B-V) = 0.11 for this galaxies, from which one obtains a reddening of 0.209 mag in $C - T_{1}$. This value, together with the slightly non-linear [Fe/H] versus $(C-T_{1})_{o}$ relation of Harris, Harris \& Geisler (2004), were used to derive the metallicities for the clusters that are listed in Table 1. It, however, should be emphasized that the Harris et al. (2004) relation between [Fe/H] and $(C-T_{1})_{o}$ was derived for very old clusters. Due to the well- known degeneracy between age and metallicity younger clusters will appear to be more metal deficient than they actually are. It is noted in passing that the cluster C150 (which is not included in Table 1 because of the absence of color information) appears to be the only heavily reddened cluster in the sample of Harris et al.(2006). Following these authors it will also be assumed that one pixel of the Advanced Camera for Surveys corresponds to a linear scale of 1.0 pc at the distance of NGC 5128, {\it i. e.} that D = 4.1 Mpc.

\subsection{Comparison between the metallicity distributions of globular clusters in the Galaxy and NGC 5128.}

The data on Mv and [Fe/H] for NGC 5128 and for the Galaxy are
plotted in Figure 1 and Figure 2, respectively. A striking difference between these two globular cluster systems is seen to be that the apparently most metal-poor clusters in NGC 5128 are systematically fainter than are the less metal deficient clusters in this galaxy. No such dependence of luminosity on metallicity is seen in the Galactic globular cluster sample. For
97 metal-rich clusters having [Fe/H] $>$ -1.80 one obtains $<M_{v} >$ = -8.20 $\pm$ 0.14, compared to $<M_{v} >$ = -7.00 $\pm$ 0.22 for 18 metal- poor clusters with [Fe/H] $<$ -1.80. Adding errors in quadrature one then obtains a luminosity difference of 1.20 $\pm$ 0.26 mag between the metal-rich and metal-poor clusters in NGC 5128.  Since the luminosity distributions in our sample are not expected to be Gaussian one should use a non-parametric test to assess the significance of the difference between the frequency distribution of the observed luminosity difference between the metal-rich and the metal-poor clusters in NGC 5128. A Kolmogorov-Smirnov test shows that there is only a 2\% probability that the 18 metal-poor clusters with [Fe/H] $<$ -1.80 in NGC 5128 were drawn from the same parent luminosity population as the 97 clusters in this galaxy that have [Fe/H] $>$ -1.80. On the other hand Figure 2 [which is based on the data in Mackey \& van den Bergh (2005)] shows no evidence for any dependence of luminosity on metallicity among Galactic globular clusters. Taken at face value the data plotted in Figure 1 would suggest that the earliest generation of globular clusters in NGC 5128 was less massive than subsequent generations of globular clusters formed in this galaxy. Alternatively, and perhaps more plausibly, one might assume that many of the clusters with ``metallicity'' [Fe/H] $<$ -1.8 are actually intermediate-age objects that appear to be metal-poor because they are relatively young and blue. If the latter interpretation is correct than the observed faintness of the clusters with ''metallicity'' [Fe/H] $<$ -1.8 is due to the fact that the intermediate-age clusters in NGC 5128 are, on average, less massive than the old globular clusters with [Fe/H] $>$ -1.8. This speculation receives some support from the observation that NGC 5128 contains two clusters which, extrapolating the non-linear relation of Harris et al. (2004), have [Fe/H] $\sim$ -4.5. Due to uncertainty of the extrapolated calibration it would probably be safer to conclude that these two clusters actually have [Fe/H] $<$ -2.5. This is lower than the value [Fe/H] = -2.29 that is obtained for NGC 5053, which is the most metal-poor Galactic globular cluster. It would be clearly be important to reobserve the clusters PFF029 ([Fe/H] = -4.20) and C119 ([Fe/H] = -4.78) to see if these objects are very old super metal-poor globular clusters, or if they are, in fact, relatively young clusters that are only mildly metal deficient. The most metal-rich cluster in the present sample is C151 which has [Fe/H] = +0.05. This is (within the limited accuracy of the data) probably not significantly different from the most metal-rich Galactic globular (Liller 1) for which [Fe/H] = +0.22.

\subsection{Luminosity-diameter relation of clusters.}

In Figure 1 clusters with FWHM $>$ 10 pc have been plotted as open circles. Inspection of this figure shows that all of the large clusters associated with NGC 5128 are faint. Figure 2 shows that a similar situation prevails in the Milky Way system where all but one of the clusters with $R_{h} >$ 10 pc (also shown as open circles) are seen to be of below average luminosity. [The lone exception is NGC 2419, which van den Bergh \& Mackey (2004) regard as the possible stripped core of a dwarf spheroidal galaxy]. It is tempting to speculate that the low luminosities of large clusters in both NGC 5128 and the Galaxy are due to the fact that these objects were formed in dwarf galaxies [which are known to contain mostly faint globulars (van den Bergh 2006)] that were later captured by NGC 5218 and the Galaxy, respectively.

\subsection{Globular clusters and dwarf spheroidal galaxies}

Mackey \& van den Bergh (2005) showed that the globular
clusters and dwarf spheroidal companions to M31 and the Galaxy appear to be separated by the empirical relation

\hspace*{5cm} Log $R_{h}$ (pc) = 0.25 $M_{v}$ + 2.95, \hspace{5cm}(1)

with dwarf spheroidals lying above (to the left) of this relation and globular clusters below (to the right) of it.
All of the objects in NGC 5128 for which Martini \& Ho (2004) list $M_{v}$ and $R_{h}$ values lie above the line defined by Eq. (1) and should therefore probably be regarded as dwarf spheroidals. Assuming that FWHM $\simeq$ 2 $R_{h}$ one would expect the globular clusters and dwarf spheroidal galaxies in NGC 5128 to be separated by the relation

\hspace*{5cm}log FWHM (pc) = 0.25 $M_{v}$ + 3.25.  \hspace{4cm}(2)                        

A plot of Mv versus FWHM for all of the clusters in Table 1 is shown in Figure 3. This Figure shows that the objects C3, C4, C6, C7, C18, and C30 in NGC 5128 fall above (or to the left) of the line defined by Eqn. 2. However, it is interesting to note that this figure does not appear to show a clear-cut segregation between putative dwarf spheroidals and the normal globular clusters.
This result may indicate that the boundary between dwarf spheroidal galaxies and globular clusters differs in position (or in its nature) for the companions to M31 and NGC 5128. Some of this observed difference might be due to stronger tidal stresses in the NGC 5128 environment.

\section{CONCLUSIONS}

Inter comparison of the M$_{v}$ versus [Fe/H] diagrams of the globular clusters surrounding the Galaxy (Mackey \& van den Bergh 2005) and of the putative globulars surrounding NGC 5128 shows an interesting difference. The luminosities of Galactic globular clusters appear to be independent of their metallicities, whereas NGC 5128 appears to exhibits a lack of luminous metal-poor ([Fe/H] $<$ -1.8) globular clusters that is significant at the 98\% level. The most plausible explanation for this difference is that a significant fraction of the apparently metal-poor globulars in NGC 5128 are actually intermediate-age clusters that appear to be very metal-poor because they are relatively young and blue.  On this hypothesis the two apparently super metal-poor clusters in this galaxy would actually be relatively young objects.
If these speculations are correct then the observed luminosity differences indicates that intermediate-age clusters in NGC
5128 are typically less massive than the old globulars associated with this galaxy.

In both the Galactic globular cluster system, and in the
cluster system surrounding NGC 5128, the largest globular clusters are found to have below average luminosities. It is well known (van den Bergh 2006) that the globular clusters in dwarf galaxies are, on average, less luminous than those associated with giants.  Hence it appears likely that the low luminosity of the most extended globulars (which mainly occur at large galactocentric distances) is due to the fact that many of these objects were initially formed in dwarf companions that were subsequently captured by the Galaxy or by NGC 5128.

Finally, in a size versus luminosity plot, the companions to NGC 5128 do not appear to exhibit the clear-cut dichotomy between globular clusters and dwarf spheroidals that is observed (Mackey \& van den Bergh 2005) among the companions to the Galaxy and M31. It is tentatively suggested that this difference might, at least in part, be due to stronger tidal stressing of dwarf spheroidals in the neighborhood of NGC 5128.

It is a pleasure to thank Christopher Conselice, Bill Harris
Dougal Mackey, Alan McConnachie and Eric Peng for their very helpful comments on an early draft of the present paper. I am also indebted to a particularly helpful referee who reminded me forcefully of the possible effects of the age-metallicity degeneracy on the interpretation of the cluster color data. Finally, thanks are also due to Brenda Parrish and Michael Peddle for technical support.

\begin{deluxetable}{lrrr}
\tablewidth{0pt}   
\tablecaption{Absolute magnitudes and FWHM diameters for globular clusters in NGC 5128}
\tablehead{\colhead{Name}  & \colhead{M$_{v}$}  &\colhead{[Fe/H]} & \colhead{FWHM}\\  &\colhead{mag} &\colhead{dex} &\colhead{pc}}

\startdata

PFF011  &   -9.20  &  -1.57  & 4.13 \\
C29     &  -10.22  &  -0.43  & 2.92 \\
PFF016  &   -8.42  &  -0.56  & 4.36\\
PFF021  &   -8.94  &  -1.78  & 3.14\\
C30     &  -11.07  &  -0.68  & 5.71\\
PFF023    & -8.78  &  -1.27  & 3.70\\
AAT111563 & -7.70  &  -2.46  & 4.20\\
C6    &   -10.67   & -0.41   & 3.94\\
PFF029  &   -8.65  &  -4.20  & 6.08\\
C4     &   -10.25  &  -1.43  & 6.10\\
PFF031  &   -8.80  &  -1.35  & 3.37\\
PFF034  &   -8.37  &  -1.29  & 4.31\\
C32     &   -9.90  &  -0.31  & 2.99\\
PFF035  &   -8.62  &  -0.31  & 4.70\\
C43    &    -9.68  &  -1.24  & 4.69\\
C12    &   -10.39  &  -0.34  & 4.15\\
WHH09   &   -9.45  &  -1.11  & 4.63\\
f2.GC70 &   -7.71  &  -1.96  & 7.21\\
f2.GC69 &   -8.45  &  -0.63  & 2.87\\
AAT113992 & -7.93  &  -0.39  & 4.85\\
C14    &   -10.34  &  -0.95  & 4.38\\
PFF041  &   -8.74  &  -1.38  & 3.40\\
ATT115339 & -8.19  &  -1.19  & 3.67\\
C3     &   -11.24  &  -0.55  & 5.61\\
PFF052  &   -8.33  &  -1.45  & 3.59\\
WHH16/K102   & -9.15   & -0.26  & 3.60\\
f2.GC31  &  -7.60  &  -1.58  & 4.02\\
f2.GC28  &  -6.59  &  -0.85 &  7.40\\
ATT117287 &   -7.30 &   -1.63 &  6.87\\
f2.GC23  &  -9.56   & -0.84  & 3.22\\
f2.GC20  &  -6.74   & -0.52  & 5.46\\
K131    &   -9.01   & -0.18 &  2.76\\
WHH22   &   -9.72   & -0.99  & 3.50\\
f2.GC14  &   -6.93  &  -0.90  & 4.06\\
AAT118198  &  -8.72 &   -0.18 &  2.79\\
PFF059   &  -8.27   & -0.41  & 4.36\\
C18/K163  &  -10.86 &   -1.06 &  4.79\\
f2.GC03   & -8.37   & -2.00  & 3.33\\
ATT119508  &   -7.89  &  -0.60 &  6.46\\
C19   &    -10.20  &  -0.93  & 4.64\\
PFF063  &   -8.30  &  -2.51  & 3.15\\
G284   &    -8.34  &  -0.56 &  2.85\\
PFF066  &   -8.31  &  -0.99  & 3.89\\
ATT120336  &   -8.08 &   -0.63 &  4.74\\
ATT120976  & -7.76 &   -1.26  & 4.19\\
PFF079  &  -8.61  &  -1.34  & 3.05\\
C104   &    -8.35 &   -1.58 &  3.58\\
G221    &   -8.92  &  -0.82  & 3.75\\
PFF083  &   -8.31  &  -0.89  & 3.17\\
C25    &    -9.78  &  -0.39  & 3.44\\
G293   &    -9.06  &  -1.69 &  3.28\\
C105   &    -5.99  &  -0.44 & 10.38\\
f1.GC20  &  -6.53  &  -0.58 & 10.39\\
C7   &     -11.11  &  -1.22 &  5.83\\
G170  &     -9.02  &  -0.58 &  3.72\\
C36    &    -9.81  &  -1.62 &  3.57\\
f1.GC15  &  -8.22  &  -0.09 &  3.36\\
f1.GC14 &   -8.08  & -1.50  & 6.95\\
f1.GC34 &   -6.79 &   -0.49  & 3.15\\
f1.GC21 &   -6.39 &   -0.22  & 6.83\\
C37    &    -9.79 &   -0.88 &  3.15\\
C111  &     -6.39 &   -2.20 &  4.33\\
C112   &    -6.44  &  -0.92 &  9.79\\
C113   &    -8.63  &  -1.62 &  3.30\\
C114   &    -6.33  &  -0.26 & 12.96\\
C115  &     -8.24  &  -1.64  & 5.62\\
C116  &     -5.91  &  -0.37 &  5.34\\
C117  &     -8.49  &  -0.30 &  3.75\\
C118  &     -7.08  &  -0.44 &  3.98\\
C119  &     -7.08  &  -4.78 &  7.09\\
C120  &     -6.32  &  -0.60 & 15.64\\
C121  &     -5.30  &  -2.49 &  5.79\\
C123  &     -7.51  &  -0.98 &  4.05\\
C124  &     -6.18  &  -1.30  & 6.61\\
C125  &     -7.09  &  -0.91  & 6.07\\
C126   &    -5.09  &  -1.69  & 4.08\\
C127   &    -6.13  &  -1.08  & 6.13\\
C128   &    -6.80  &  -0.06  &10.23\\
C129  &     -6.92  &  -0.74  & 4.45\\
C130  &     -7.90  &  -1.64  & 3.69\\
C131 &     -7.55   & -1.66   &5.87\\
C132   &    -8.85  &  -1.35  & 3.60\\
C134    &   -7.04  &  -0.79  & 6.11\\
C136     &  -6.56  &  -1.84  & 4.90\\
C138  &     -7.78  &  -1.17  & 3.80\\
C139    &    -8.89 &   -0.54 &  3.26\\
C140  &     -7.92  &  -1.13  & 5.01\\
C141   &    -6.80  &  -1.86  &12.11\\
C143   &    -7.22  &  -1.71  & 5.83\\
C144   &    -5.79  &  -1.95  & 8.76\\
C145   &    -9.94  &  -1.27  & 3.99\\
C146   &    -7.81  &  -0.71  & 4.08\\
C147   &    -7.70  &  -1.24 &  5.77\\
C148   &    -7.54  &  -2.02 & 10.55\\
C149   &    -8.07  &  -1.71 &  3.67\\
C151   &    -7.80  &  +0.05 &  3.42\\
C155   &    -6.39   & -1.00 &  9.20\\
C156   &   -10.10  &  -0.18 &  3.66\\
C158   &    -7.69 &   -1.05 &  7.41\\
C159   &    -7.82 &   -0.34 &  3.30\\
C161   &    -8.49 &   -0.39 &  3.28\\
C162   &    -6.84 &   -2.60 &  9.15\\
C163   &    -7.32 &   -1.18 &  8.06\\
C165   &    -9.58 &   -0.42 &  3.71\\
C169   &    -7.36 &   -2.08 &  6.27\\
C170   &    -5.59 &   -2.28 &  4.36\\
C171   &    -7.08  &  -1.12 & 11.31\\
C172   &    -6.84  &  -1.99 &  5.76\\
C173   &    -6.69  &  -0.19 &  9.40\\
C174   &    -6.33  &  -0.63 &  4.66\\
C175   &    -5.91  &  -0.92 &  5.81\\
C176   &    -6.45  &  -2.58 &  5.32\\
C177   &    -6.66  &  -0.53 &  7.01\\
C178   &    -6.22  &  -1.35 &  7.71\\
C179   &    -6.71  &  -2.47 &  8.24\\

\enddata
\end{deluxetable}

\begin{figure}
\caption{Luminosity versus metallicity for 115 clusters associated with NGC 5128. Metal-poor clusters with [Fe/H] $<$ -1.8 are, on average, less luminous than are the more metal-rich clusters. Clusters with FWHM $>$ 10 pc (plotted as open circles) are seen to be significantly fainter than the smaller clusters which are shown as dots.}
\end{figure}

\begin{figure}
\caption{Luminosity versus metallicity for Galactic globular clusters. The data, taken from the updated version of Harris (1996), show no dependence of cluster luminosity on metallicity.  Large clusters with R$_{h} >$ 10 pc (which are plotted as open circles) are seen to be systematically fainter than more compact Galactic clusters.}
\end{figure}

\begin{figure}
\caption{ Full width half maximum (FWHM) cluster diameter versus cluster luminosity for the NGC 5128 cluster system. The plotted line is Eqn. (2). The data show no clear-cut segregation between globular clusters and dwarf spheroidal galaxy. In this respect NGC 5128 appears to differ from M31 and the Galaxy.}
\end{figure}


\begin{references}

\reference{ } Geisler, D. \& Forte, J. C., 1990, ApJ, 350, L5 

\reference{ } Harris, W. E. 2001 in Star Clusters, Eds L. Labhardt \&B. Binggeli, Saas Fee Advanced Course 28 (New York: Springer)

\reference{ } Harris, W. E.,Harris, G. L. H., Barmby, P., McLaughlin, D. E. \& Forbes, D. A. 2006, =astro-ph/0607269

\reference{ } Harris, G. L. H., Harris, W. E., \& Geisler, D. 2004,AJ, 128, 723

\reference{ } Mackey, A. D., \& van den Bergh, S. 2005, MNRAS, 360, 631 

\reference{ } Martini, P. \& Ho, L. C. 2004, ApJ, 610, 233 

\reference{ }van den Bergh, S. 2006, AJ, 131, 304 

\reference{ }van den Bergh, S. \& Mackey, A.D. 2004, MNRAS, 354, 713

\end{references}
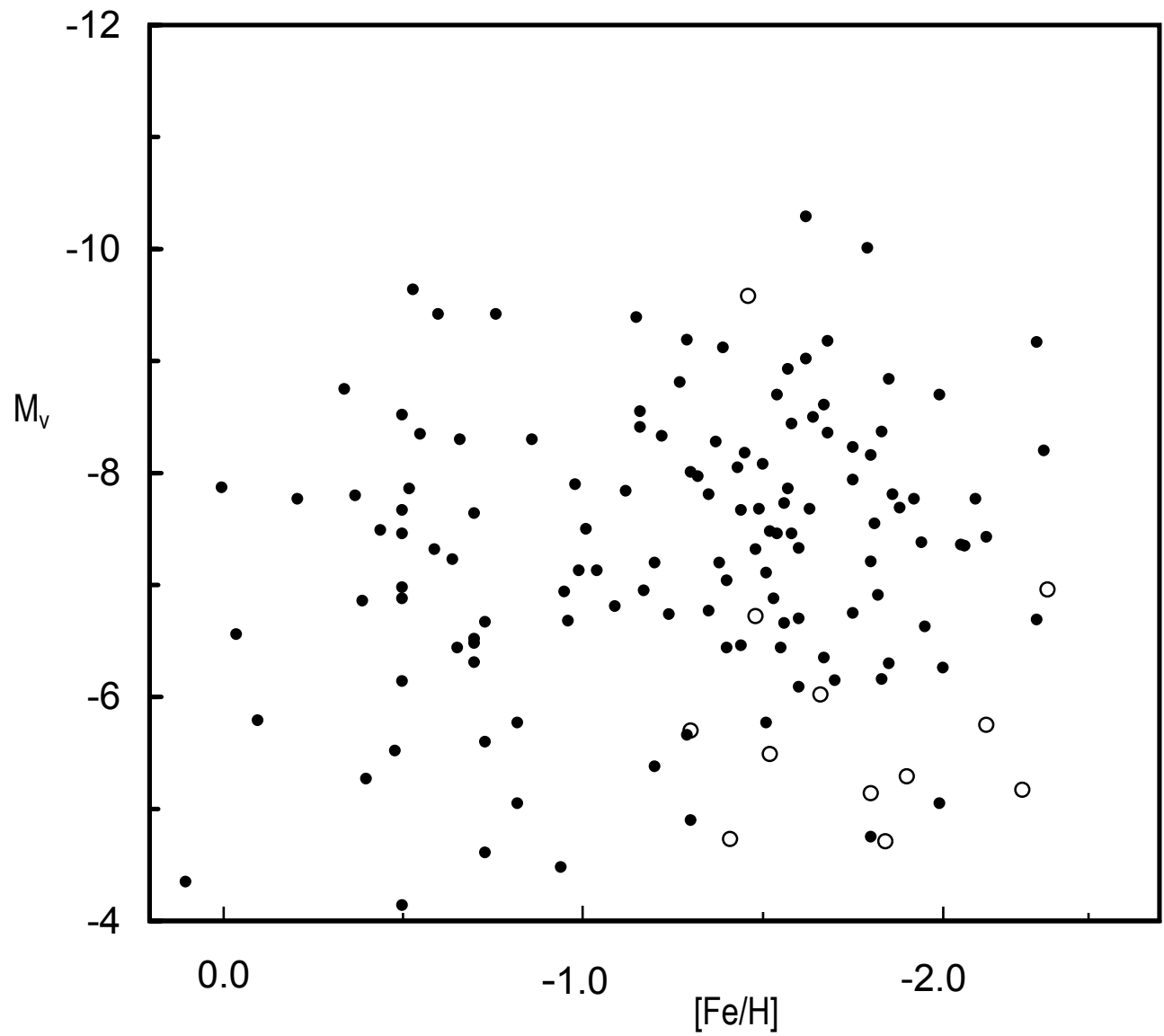
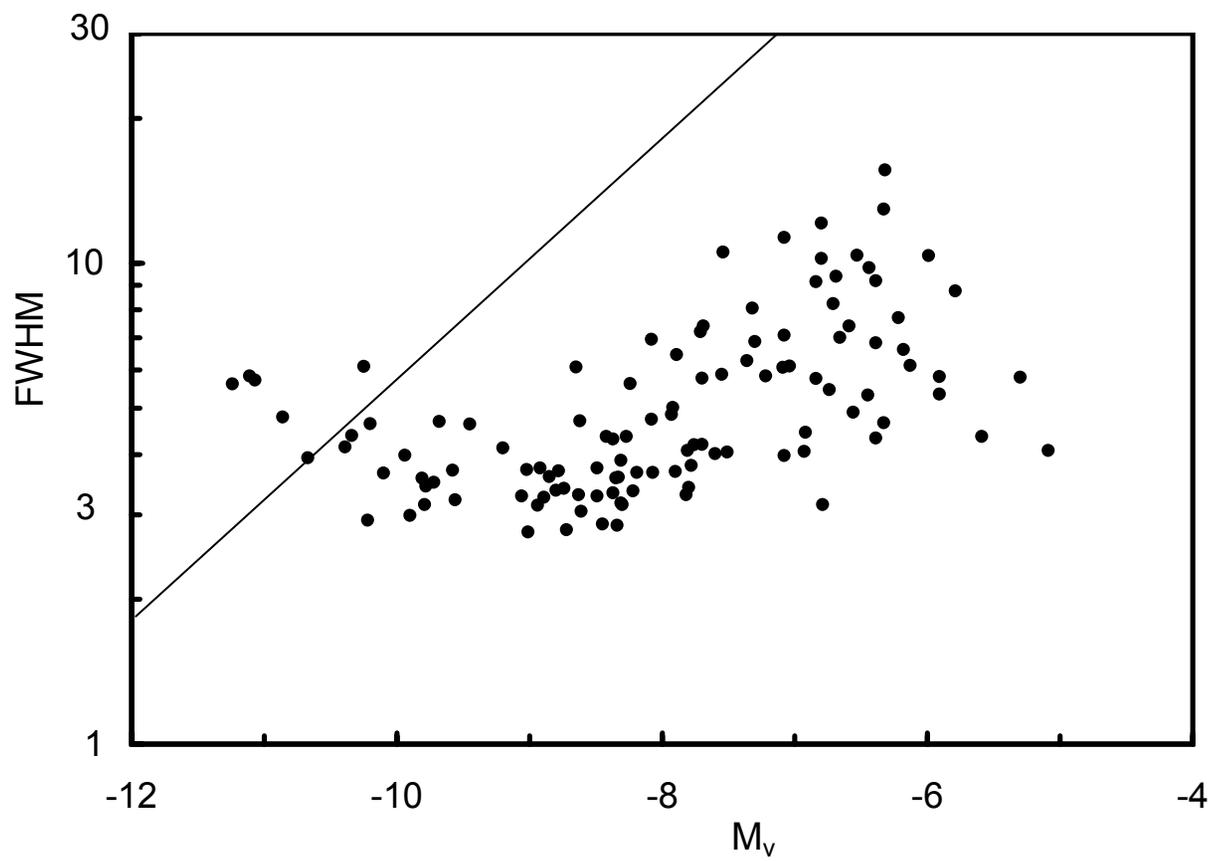
\end{document}